This is the author-accepted version of the paper published in the 2026 IEEE Conference on Artificial Intelligence (CAI), Granada, Spain, 8–10 May 2026.

# AI-Powered Symptom Assessment and User Experience: A Case Study of Simtomi and Simtomi-Care

Jinha Lee
*DeVoe Division of Business*
*Indiana Wesleyan University*
Marion, USA
jinha.lee@indwes.edu

Chan Hyung Lee
*Research Institute of Mediark*
*Mediark*
Seoul, Republic of Korea
judah@mediark.io

Hyunsung Lee
*Research Institute of Mediark*
*Mediark*
Seoul, Republic of Korea
ben@mediark.io

Seunghwan Kim
*College of Medicine,*
*Ewha Womans University*
Seoul, Republic of Korea
kimshwan901@gmail.com

Ban Hyung Lee
*Research Institute of Mediark*
*Mediark*
Seoul, Republic of Korea
johann@mediark.io

Minjun Shin
*Department of Biology*
*Indiana University*
Bloomington, USA
paulshin@iu.edu

Hojin Shin
*Hamilton Southeastern HS*
*Indiana Wesleyan University*
Fishers, USA
hojinshin2027@gmail.com

Jungdo Park
*Research Institute of Mediark*
*Mediark*
Seoul, Republic of Korea
jason@mediark.io

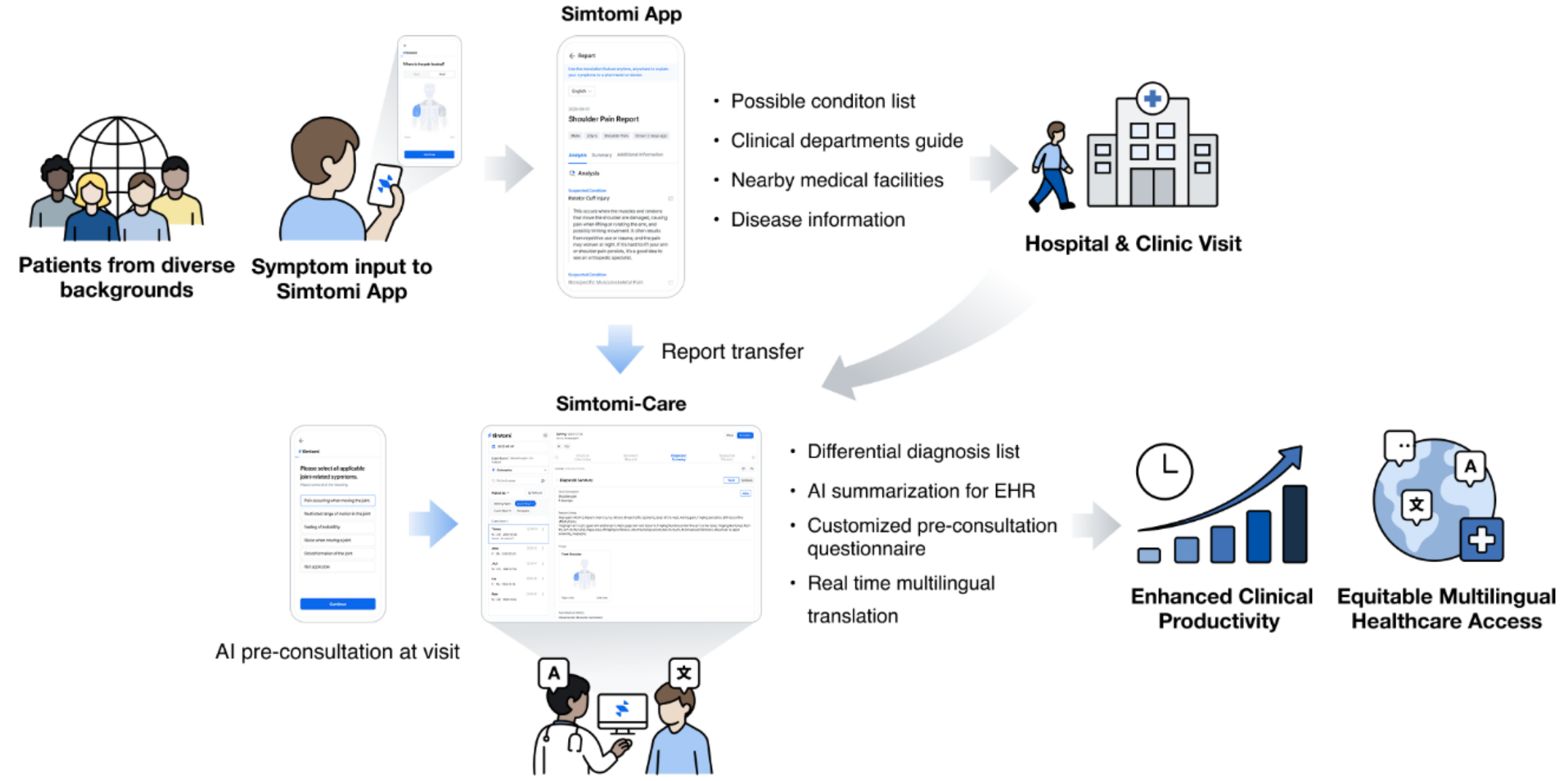

Fig. 1. Workflow of the Simtomi system

***Abstract*— Digital symptom checkers are widely used for quick guidance on health concerns, yet many systems still face challenges in collecting accurate information, supporting communication, or integrating with clinical workflows. To explore how these tools function in real use, we examine the case of the Simtomi system, which pairs a multilingual symptom assessment application with a provider-facing platform. Empirical studies were conducted in two countries. In South Korea, based on participants' firsthand experience, we found that the system improved how patients communicated their symptoms and helped clinicians review cases more efficiently through structured summaries aligned with diagnostic reasoning. In the United States, responses from prospective users and healthcare professionals highlighted the value of multilingual support, structured questioning, and the system's potential to assist clinical coordination. These findings offer a grounded account of how AI-based symptom assessment tools can operate across different healthcare contexts and provide broader insight into usability, trust, and usefulness in digital health.**

## I. Introduction

Digital symptom checkers have emerged as widely used entry points for people seeking health information online. Their popularity reflects both increasing demands on health systems and the desire of patients to obtain rapid guidance without the time and cost barriers of in-person care. However, systematic reviews highlight inconsistent performance, with ongoing concerns about accuracy and reliability [1], [2], [3], [4]. Even when technically sound, many symptom checkers constrain input formats, overlook contextual details, and provide limited follow-up guidance, resulting in incomplete patient journeys [5], [6]. The findings highlight that technical performance alone is insufficient: patient-facing tools must also be usable and trustworthy. Design studies show that explanation style and interface modality directly shape users' perceptions of credibility, authority, and comfort [7], [8]. Recent work further emphasizes that emotional reassurance and cultural responsiveness are as important to acceptance as diagnostic outputs [9].

To advance these conversations, we examine the case of Simtomi, a multilingual, AI-powered symptom checker paired with a provider-facing platform. Simtomi is an actively deployed AI-based symptom assessment system developed and operated by Mediark Inc.[10], a digital healthcare company in South Korea. Unlike tools that stop at basic triage, Simtomi is designed as a "symptom-to-system bridge" that supports both patients and clinicians. Its mobile application allows users to report symptoms in their preferred language and receive guided questioning, while its clinical module generates structured summaries for integration into electronic health records (EHRs). By combining multilingual accessibility with provider-oriented outputs, Simtomi offers a compelling case study for investigating whether design features can reduce communication barriers and align with goals of improving the reach of digital health systems (Fig. 1). Thus, this study aims to examine how the system may facilitate access to care and improve communication between patients and clinicians, and how users and healthcare professionals evaluate the usability and usefulness of the integrated Simtomi system. To address these aims, our approach draws on insights gathered from two different healthcare settings. Research on digital health argues that design interventions must be evaluated in diverse contexts, as structural differences in health systems and cultural practices shape both uptake and trust [11], [12], [13]. We therefore conducted a two-phase case study of Simtomi across South Korea and the United States. Study Phase 1 involved in-depth interviews with Korean users and physicians to capture perceptions of usability and clinical alignment. Study Phase 2 scaled this inquiry through a U.S.-based survey of users, analyzed with sentence-level semantic clustering[14], to surface broader patterns of usability, trust, and usefulness. This multi-phase, mixed-methods approach allowed us to integrate close interpretation with computational analysis and to compare perspectives across cultural and institutional contexts.

## II. BACKGROUND

### A. *Simtomi System*

The Simtomi system consists of two interlinked modules: the Simtomi mobile application for patients and Simtomi-Care for providers and clinical environments. The mobile app allows patients to report symptoms in their preferred language through short free-text input, after which the system guides them through an AI-driven questioning process that suggests probable conditions and recommends appropriate medical departments or nearby healthcare facilities using location-based services (Fig. 2). In addition, a pharmacy support module with integrated translation allows patients to communicate simple symptoms and request medications across languages, reducing barriers in pharmacy encounters for international patients. Simtomi-Care extends the system into clinical settings as a provider-facing platform. Patient information, entered either via the mobile app or directly at the point of care, is processed by an AI-driven pre-consultation questionnaire module. This module converts symptom descriptions into medically precise terminology, generating structured summaries for clinical use (Figure 3). The summaries are delivered to physicians via the Simtomi-Care program installed on clinic workstations, where they can be reviewed and integrated into existing EHR systems. Simtomi-Care supports workflow efficiency by reducing documentation workload and integrating with hospital platforms. It also provides differential diagnosis guidance and multilingual communication tools, helping providers deliver more streamlined and multilingual care. Simtomi has achieved ISO 13485 certification for medical device quality management, ISO 27001 for information security, and ISO 17100 for translation services. These certifications indicate adherence to global benchmarks of safety, security, and reliability in healthcare deployment. The app, available for use internationally, enhances accessibility for multilingual populations across countries by supporting 14 languages, and its integration with clinical workflows is currently in use in South Korea, Vietnam, and South Africa.

### B. *Technical Implementation of Simtomi*

The development of the Simtomi system involved collaboration across technical and clinical domains, with contributions from engineers, physicians, departmental specialists, and multilingual translators. These groups worked together on database construction and labeling, and applied appropriate AI techniques at each stage of the process, such as symptom entry, medical questioning, disease inference, summarization, and translation (Fig. 4).

*NLP-Based Chief Complaint Matching.* At the intake stage, a multilingual corpus of symptom expressions enables patients to describe their conditions through short free-text input or by selecting associative terms. The input is then processed by a natural language processing (NLP) module, which semantically maps patient descriptions to standardized chief complaints [10], [15]. This design lowers barriers for patients from diverse linguistic backgrounds and improves accuracy of chief complaint selection, which is critical for subsequent processing of symptom data collection. In Simtomi, 14 languages are supported (e.g., Korean, English, Spanish, Chinese, and Arabic), and the multilingual corpus of associative symptom expressions was constructed and validated by a professional multilingual translation team.

*Machine Learning–Based Adaptive Questioning & Disease Inference Models.* During the adaptive questioning stage, diagnostic questions are dynamically generated based on symptom nodes derived from user responses, age, and gender. A machine learning–based predictive model continuously refines the probability of candidate conditions and generates targeted follow-up questions within a loop structure [16], [17]. Symptom- and disease-related diagnostic questions were developed from standard medical textbooks and clinical guidelines, followed by multi-stage review with 30 departmental specialists. Based on feedback from physicians and patients in clinical settings using the Simtomi system, and further refined through iterative testing, the questioning sequence was limited to fewer than 20 items to minimize user burden while maintaining diagnostic coverage. The information nodes collected during adaptive questioning are classified into symptoms, symptom-specific details, risk factors, and contextual factors. Each node can influence multiple diseases, creating a complex network (Figure 5). As prior studies evaluate diagnostic reasoning using Top-1, Top-3, or Top-5 accuracy metrics [16], [18], the Mediark developers assigned baseline weights to medically significant symptoms and risk factors, and refined them at periodic intervals through feedback-driven machine learning using clinical data validated by medical experts. In a preliminary internal evaluation, 20 representative cases per specialty were presented to 2–3 physicians in each department. The correct diagnosis was identified in 62% of Top-1 outputs, 76% of Top-3 outputs, and approximately 88% of Top-5 outputs (excluding pediatrics). For context, prior studies report average Top-1 accuracies of 34–51% and Top-3 accuracies of 51–71% across widely used symptom checkers [16], [18], with the highest-performing systems approaching 70% at Top-3 [3]. Given these comparatively strong results, formal external validation using standardized clinical vignettes remains a priority for future work. The system presents patients or physicians with a ranked list of potential diseases, and in the Simtomi application, the output includes not only the disease names but also detailed information and guidance related to each condition.

*LLM-Driven EHR-Compatible Summarization of Patient Inputs.* Recent research shows the use of LLMs to convert unstructured clinical text into structured representations [19]. Simtomi similarly applies large language models (LLMs) to convert patient interaction data (symptom entries, chat transcripts, etc.) into structured summaries that integrate with EHRs. The system has made deliberate efforts to adopt widely accepted summarization practices, guided by feedback from practicing clinicians. For example, patient narratives such as "pain in the right upper side of the belly" are a utomatically converted into standardized clinical terms such as "RUQ pain." The team fine-tuned key medical terms across languages using professional translators and prompt-engineering strategies. To minimize the possibility of hallucinations and other inaccuracies, the developers closely aligned each summarized output with the original patient input and refined the mapping through repeated testing.

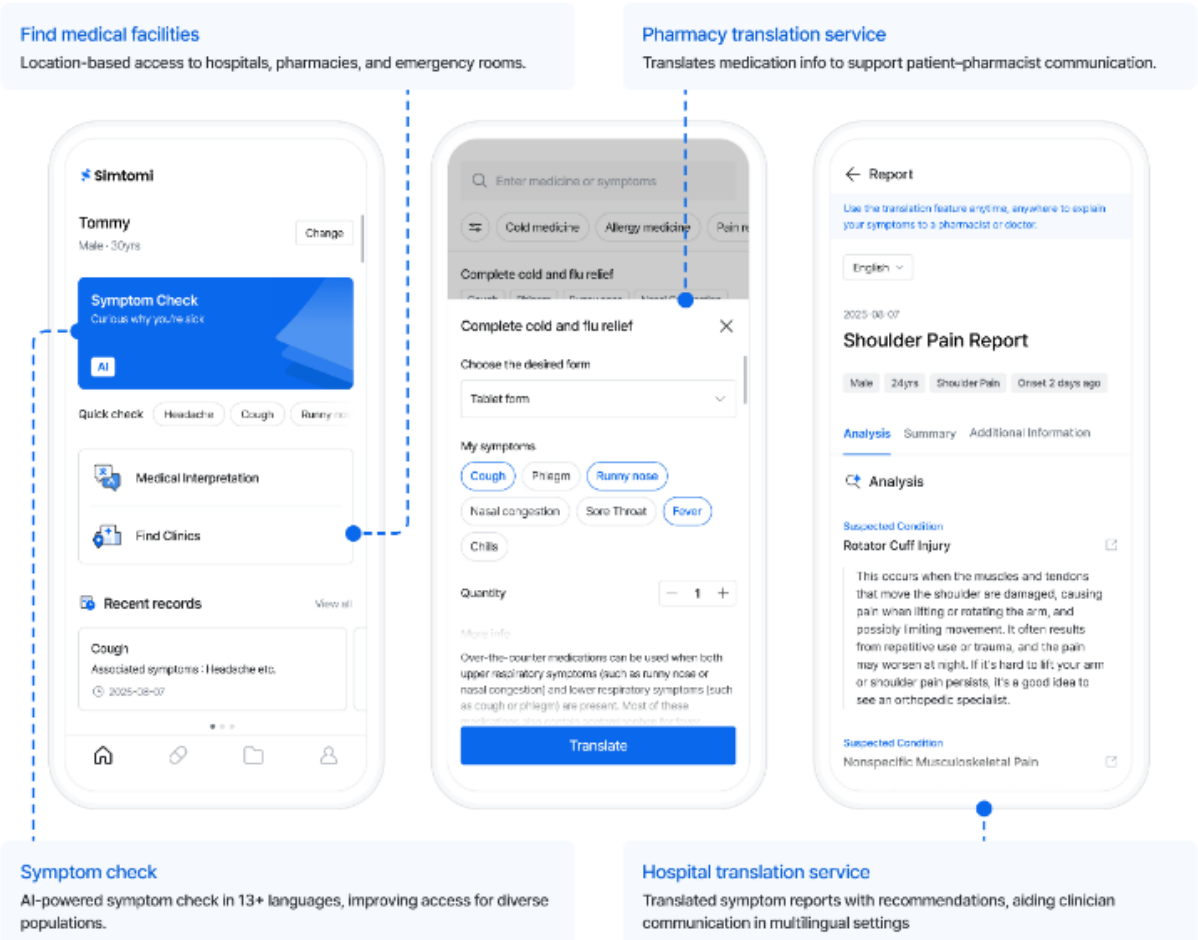

Fig. 2. Main interface and key functions of the Simtomi mobile app.

## Methodology

### C. Study Phase 1: South Korea

Study Phase 1 was carried out in South Korea to capture perspectives from two groups: users who had direct experience with Simtomi, and physicians who had used the integrated Simtomi system in clinical contexts. We recruited participants through professional networks, local clinics, and user communities. Users were individuals who had personally experienced Simtomi for their own or family health concerns, while health professionals were physicians who used the integrated system in clinical practice across a range of specialties. In total, 11 consumer participants and four physicians took part. The consumer group had an average age of 34 years (four females), and the professional group an average age of 36 years (two females) (see Appendix A.3.1 for details). All participants provided informed consent. Semi-structured interviews using open-ended questions were conducted, either in person or through secure video calls. Each lasted between 20 and 30 minutes. Protocols were developed for users and health professionals. We structured our protocol around five areas of evaluation (Appendix A.1). In detail, the interview questions for users focused on their experience with the Simtomi app, while those for healthcare professionals addressed both the Simtomi app and Simtomi-Care, which they use together in practice. Consequently, the professional protocol was more targeted. Trust questions incorporated accuracy, which professionals assessed in relation to their medical diagnoses, and the usefulness questions distinguished between patient and provider perspectives.

All interviews were transcribed verbatim and coded in NVivo 12. We followed a thematic analysis process [21] with iterative rounds of open and axial coding. Codes were first developed inductively, capturing concepts that emerged from the data. Coding and category refinement were discussed among the research team to maintain consistency and analytic rigor, which resulted in an agreed structure under the five evaluation categories. Coding followed iterative researcher consensus, consistent with reflexive thematic analysis [21]. Separate codebooks were produced for users and for professionals. This allowed role-specific themes to surface while still enabling comparison across groups.

### D. Study Phase 2: United States

Study Phase 2 aimed to examine the perceived usefulness, user experience, and underlying values of the Simtomi app in the U.S., a context characterized by substantial variation in healthcare access and system fragmentation. While a few AI-powered symptom checkers exist in the U.S., many lack fully developed features that connect users directly with healthcare providers. We assessed whether Simtomi's functions could address gaps in access to care and demonstrate the potential of integrating advanced digital technologies with healthcare services. First, we recruited general users (n=201, Appendix A.3.2) living in the U.S., via CloudResearch. The same open-ended question set from Phase 1 was used to elicit in-depth written responses. Given the practical constraints of remote participation and the need to ensure data quality, participants viewed a 3-minute Simtomi walkthrough video [22] demonstrating its navigation and features (e.g., multilingual symptom input, adaptive questioning, reports). This standardized presentation minimized distraction and avoided uncontrolled variability in access. To analyze 5,743 sentences from 1,205 responses, we used sentence-BERT (SBERT), a modification of the bidirectional encoder representations from transformers (BERT) architecture designed for generating semantically meaningful sentence embeddings [14], [23]. We used the pre-trained SBERT model (all-MiniLM-L6-v2) from the SentenceTransformers library. A general-domain SBERT model was used because the responses reflected everyday descriptions of perceptions and experiences rather than specialized clinical language. Prior to embedding, responses underwent preprocessing: all text was cleaned and lowercased, with standard English stop words removed along with a small set of domain-specific terms (e.g., the app's name, generic descriptors, and common opinion verbs) to reduce noise. Each sentence was encoded as a 384- dimensional vector, and cosine similarity was used to compare responses. We then applied k-means clustering to group semantically similar responses. To determine k, we compared standard clustering validity metrics and selected the final value based primarily on Silhouette performance, considering balanced interpretability and cluster compactness. To enhance interpretability, we generated theme labels using the bidirectional and auto-regressive transformers (BART) [24], an abstractive summarization model that captures full context and produces concise, coherent labels more reliably than general-purpose LLMs. We prompted BART with centroid-nearest responses from each cluster, yielding theme names that closely reflected participant language.

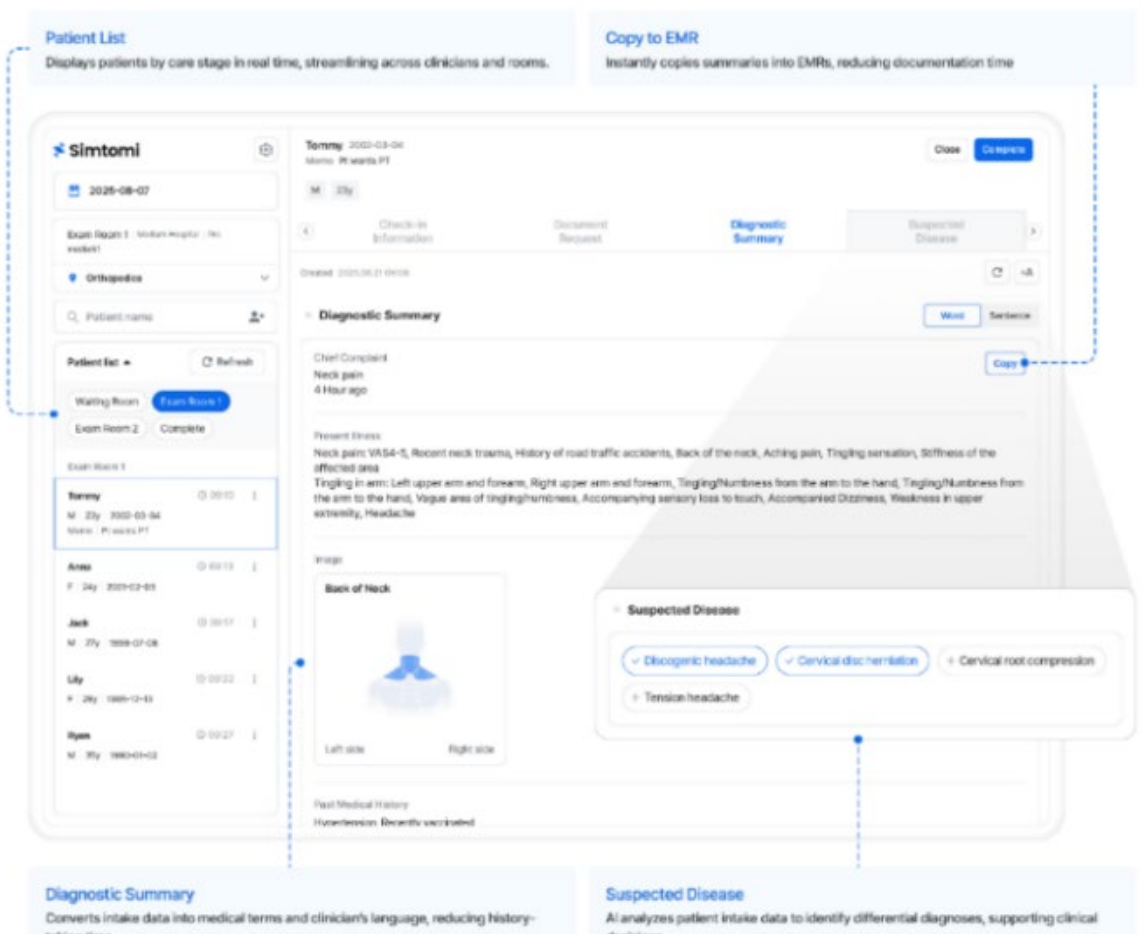

Fig. 3. Simtomi-Care provider-facing desktop application

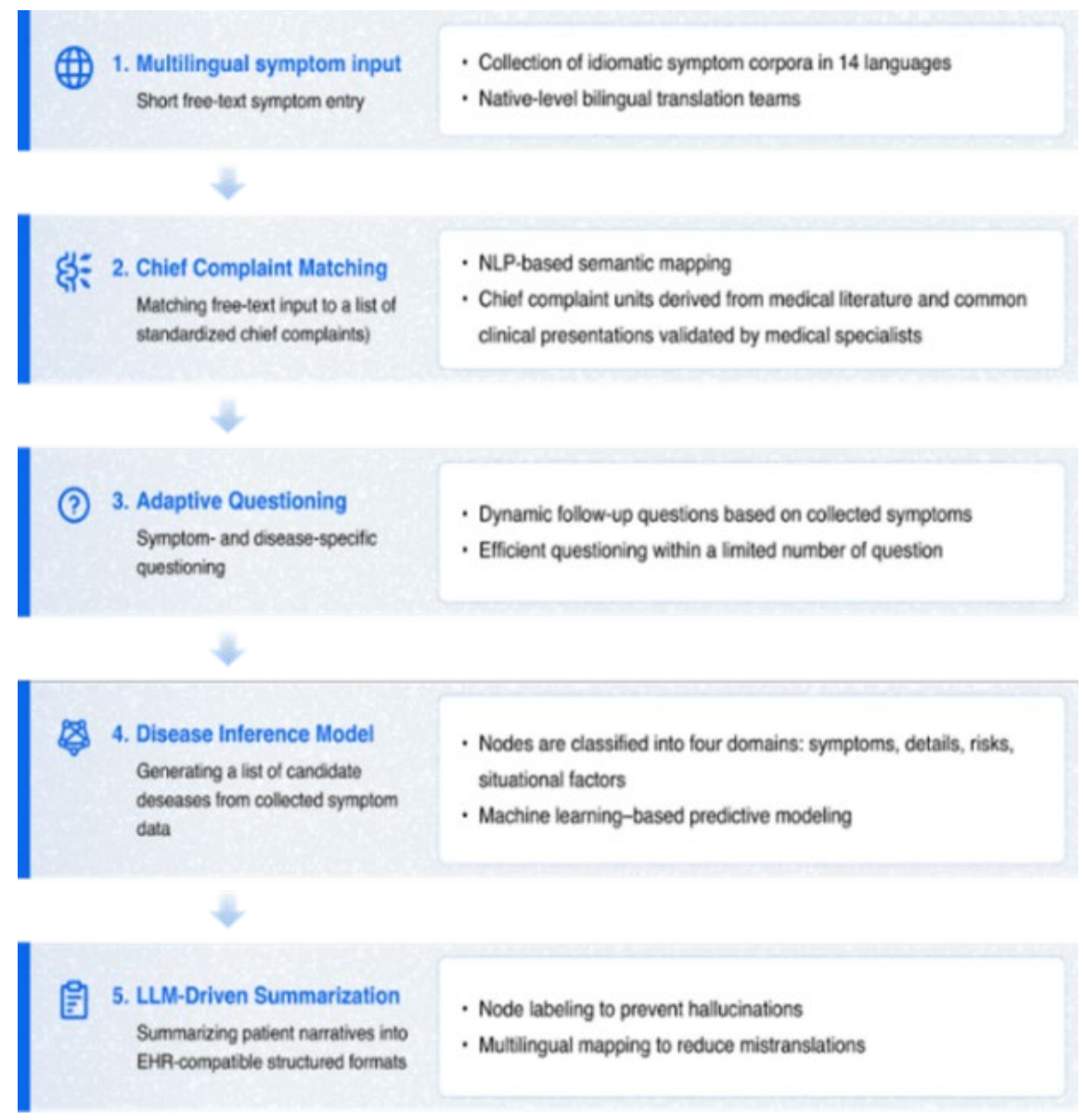

Fig. 4. Simtomi system development pipeline

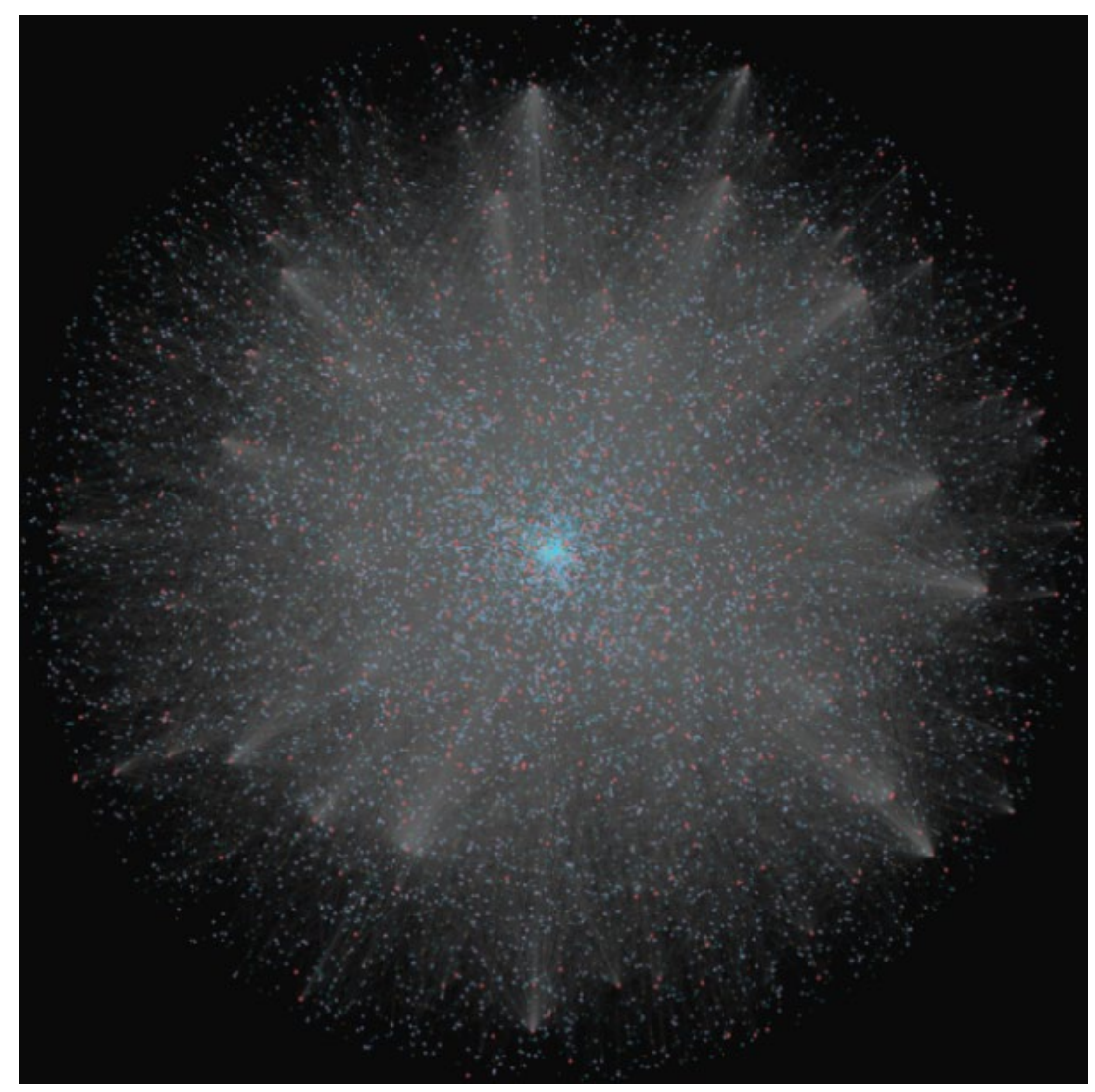

Fig. 5. Network structure of Simtomi's disease data (red) and related clinical/contextual factors. Major symptoms (sky blue), symptom details (purple), risk factors (orange), and causes or contexts (yellow) connect to disease nodes, illustrating the complexity of the medical knowledge graph.

Next, we recruited healthcare professionals through professional contacts and institutional networks. Semi-structured interviews were carried out online and averaged about 30 minutes. The protocol followed the same structure as in earlier phase. Seven professionals were interviewed: three female and four male participants, including three physicians, three nurse practitioners, and one nurse, with an average of 16 years of clinical experience across various specialties (internal medicine, family medicine, emergency medicine, pediatrics; Appendix A.3.3). Participants were asked to install and test the Simtomi app prior to the interview by entering hypothetical symptoms, ensuring direct firsthand use of the patient-facing tool. At the beginning of the session, they also viewed a short demonstration of Simtomi-Care [25], which illustrated how patient-entered data could transfer into a provider-facing workflow. While the app is available for download in the U.S., Simtomi-Care is not currently deployed, so participants' reflections on the provider module were informed by the demonstration. All interviews were transcribed verbatim and analyzed using NVivo 12. Open codes were inductively generated and grouped into axial categories under the five domains. Coding was conducted iteratively by the research team, reaching agreement through discussion.

## III. Results

### A. Results of Study Phase 1

Users in South Korea described Simtomi's visual design as both professional and approachable. One participant noted, "It was clean and simple, listing only the important information. Overall, the UI felt like it was tailored to users." (LK2) In terms of usability, participants consistently highlighted the ease of navigation and the guiding role of the step-by-step flow. Many appreciated the way symptom prompts encouraged them to reconsider their conditions, as one user shared: "As I answered the questions, it made me re-check my symptoms and think more carefully about my condition." (UK11) These reflections suggest that despite minor learning curves, the interaction flow was perceived as accessible even by those less experienced with technology. Trust emerged as a salient dimension of user experience. Several participants attributed trust to perceived source credibility, noting that physicians were involved in developing the system and that this raised their confidence. As one participant explained, "The founder is a doctor, so I felt the product itself carried expertise, and that made me trust it more." (UK4) Participants also felt reassured when their inputs were reframed in professional medical language. Trust was further reinforced when the system's outputs aligned with clinical encounters: "I used it late at night, and the next day the doctor told me almost the same thing. That really increased my trust." (UK7) When reflecting on usefulness, participants emphasized the app's value in decision-making and communication. Several explained that the reports helped them decide whether immediate medical care was necessary, with one saying, "The report helped me decide whether to go to the hospital right away or wait—it gave me clarity." (UK7) Others described how it supported communication with physicians, making explanations more accurate and less stressful: "It helped me explain my symptoms to the doctor more clearly and confidently." (UK3) Beyond saving time, the tool was perceived as giving users a stronger role in directing their own care. In sum, users perceived the tool as extending beyond usability into a form of health agency, suggesting its broader potential in contexts where patients often feel disempowered. Appendix A.2.1 provides detailed illustrative interview quotes.

Physicians in South Korea emphasized visual design as a contributor to trust and usability. The interface was seen as modern, simple, and intuitively clear. With respect to usability, physicians valued how the flow of questioning mirrored medical reasoning. One remarked, "The way patients answered step by step, almost like a checklist, matched well with the consultation process." (PK5) For many, these design elements not only improved usability but also integrated seamlessly into clinical workflows. Trust was strongly reinforced by the system's capacity to transfer patient inputs into clinically accurate language. A physician noted, "The questions and summaries felt like something a doctor would write, using terms we commonly use."(PK2) Others highlighted the breadth of diagnostic coverage, appreciating suggestions outside their immediate specialty.. Furthermore, the professional group described usefulness in terms of clinical plausibility, efficiency, and communication support. They evaluated the suggested diagnoses to be clinically sound and relevant to routine practice: "Most of the diagnoses it suggested were reasonable ones we would expect to see in practice." (PK2) The app was also seen as saving time by streamlining patient presentations: "It probably shortened the first consultation by about two minutes." (PK3) Taken together, professionals trusted the system when its questioning style aligned with diagnostic logic, and they viewed the structured summaries as directly translatable to medical documentation. Their reflections highlight a dual benefit: on the one hand, workflow efficiency that saved time and reduced redundancy; on the other hand, diagnostic breadth that prompted clinicians to consider possibilities they might otherwise overlook. These insights, drawn from firsthand experiences of users and professionals, provided a baseline that informed the subsequent quantitative phase.

### B. Results of Study Phase 2

From each cluster, we extracted the top 10 responses nearest to the centroid based on cosine similarity and summarized them with BART to generate preliminary themes. To qualitatively validate these themes, we reviewed the top responses alongside the BART summaries. Because the machine-generated themes were often lengthy, we further condensed them through researcher consensus to produce concise final theme names. This process ensured that the themes were both data-driven and semantically interpretable. Theme assignment confidence averaged 0.431, with 30% of responses above 0.50 cosine similarity. Mean intra-theme similarity was 0.470, indicating moderate semantic cohesion consistent with cosine similarity distributions reported in embedding benchmarks [26], [27]. Manual spot-checks of stratified samples showed over 90% conceptual fit between assigned themes and original responses, supporting face validity.

Visual design responses revealed three themes. The largest theme (43.8%, n=88) emphasized intuitive and easy-to-follow design with clear colors and fonts. A second theme (32.8%, n=66) described appreciation for the dropdown layout and ease of navigation. A third cluster (23.4%, n=47) highlighted the clean, modern layout with clear icons and an interactive body map. These themes showed a mix of simplicity and structural features shaping design appeal. Usability was described through three themes. The largest (41.3%, n=83) highlighted autocomplete suggestions paired with a body chart for symptom input. Another cluster (31.8%, n=64) focused on being led by simple, easy-to-understand questions. A third group (26.9%, n=54) described the app as accessible even for non-tech-savvy users, with a natural step-by-step flow. Together, usability responses combined specific functions with general ease of interaction. Trust perceptions centered on two themes. A majority (70.1%, n=141) described the app as trustworthy through its clinic connections and "doctor-like" questioning. The remaining group (29.9%, n=60) viewed the app as generally trustworthy because it was designed by doctors, though respondents noted they would still double-check with a healthcare provider for serious issues. In sum, trust was grounded in both the app's clinical framing and its professional origins. In usefulness, two clusters emerged. One (50.7%, n=102) emphasized clear, step-by-step guidance and practical care directions, including locating nearby clinics and pharmacies. The other (49.3%, n=99) underscored self-understanding and improved communication with doctors, with many noting that the summary report would help them describe symptoms more clearly. Taken together, these themes suggest that usefulness was expressed through both personal decision support and facilitation of clinical dialogue.

The findings from healthcare professionals in the U.S. consistently highlighted Simtomi's simplicity as its defining visual characteristic. The interface was described as clear, uncluttered, and easy to navigate, with little variation in opinion compared to earlier phases. Perceptions of usability centered on ease of use and input clarity, supported by features such as keyword search, autocomplete, body maps, and plain-language prompts that lowered barriers for patients unfamiliar with medical terms. Participants also noted areas for refinement, including occasional gaps in autocomplete recognition and the length of certain question pathways. These concerns suggested that while the interaction flow was broadly accessible, improving efficiency would enhance its value for both patients and providers. Trust was described as anchored in source credibility and clinical relevance. As one physician noted, "I believe it is trustworthy. It uses standard common textbooks as references so that helps the accuracy and clinical relevance." (PU1) Interestingly, some professionals in the U.S. tied trust to data security and personal information, partly reflecting a different emphasis from Korean professionals who focused more on performance credibility. By guiding users to think through their conditions, the app was seen as reducing ambiguity and supporting more informed dialogue with providers. As one nurse practitioner explained, "It gets people thinking specifically about how to communicate what their experiences are." (PU5) Beyond communication, professionals noted that features such as associated symptom prompts and nearby care options helped patients anticipate possible diagnoses and navigate care settings. For providers, usefulness was tied to workflow efficiency and data quality. They highlighted the value of pre-populated histories that could be transferred into EHRs, reducing repetitive documentation and freeing time for direct patient care. Additionally, professionals stressed that integration with existing EHR systems would determine whether these efficiencies were realized. Appendix A.2.2 provides detailed illustrative interview quotes.

## IV. Discussion, Limitations, & Future Work

### A. Discussion

The findings reveal that across both countries, participants emphasize the interface's visual qualities, describing it as clean, simple, and visually pleasing. This consistency suggests that minimal aesthetics can reduce cognitive load and support users with varying levels of health literacy. This aligns with broader HCI findings that simplifying visual information and interaction design lowers mental effort and improves usability [28]. When discussing visual design, participants blurred aesthetics and usability, an overlap we anticipated given design's functional elements. For analytic clarity, and consistent with prior HCI work that treats these as distinct constructs [29], [30], we analyzed them separately, focusing on participants' appraisal of the interface's visual elements in their visual design responses. Meanwhile, usability centered on how effectively the system supported information access and interaction. Korean users stressed actionable health information, multilingual support, and symptom-based prompting, while U.S. users emphasized autocomplete, body charts, and step-by-step questions that eased use for those less tech-savvy. Professionals in both countries highlighted alignment with medical workflows, efficiency, differential diagnosis, and patient autonomy. Taken together, these findings show that usability considerations for AI symptom checkers extend beyond interface simplicity to encompass informational accessibility, guided interaction, and integration into both patient and professional practices.

Health professionals in both countries viewed the system as highly credible, citing its differential diagnosis capability and physician-led development. Trust was particularly strong among South Korean providers, likely reflecting their direct experience with Simtomi-Care. U.S. professionals also expressed confidence, though they raised concerns about data security and the possibility of missed diagnoses. Most notably, the multilingual option emerged as the most valued feature among U.S. professionals, who serve highly diverse patient populations. Every professional emphasized that language options improve accessibility. Users in both countries also noted that translation reduces barriers for immigrants, international residents, and travelers. Taken together, these findings indicate that multilingual support is a consistently recognized strategy for broadening access and reducing communication barriers in AI symptom checkers. When it comes to differences between the countries, the professional groups showed differing emphases. South Korean professionals emphasized diagnostic breadth and cross-specialty coverage offered by the system, while U.S. counterparts focused more on institutional integration and liability concerns. U.S. participants further highlighted its potential benefits for underserved areas. In this sense, tools like Simtomi illustrate how digital symptom assessment systems may serve different roles depending on structural conditions, with some contexts emphasizing efficiency and others prioritizing safety, integration, or equity needs. However, participants also raised concerns. For example, a few U.S. providers noted that usability is constrained by literacy, with some patients in rural areas lacking the skills needed to use the system. Older adults may also struggle without voice input, and cultural nuances in diet and health practices are difficult to capture through standardized questioning. These findings underscore that supporting diverse users involves more than adding features such as language options; it calls for ongoing attention to literacy, aging, and cultural variation in system design. U.S. professionals suggested that AI-based assessment tools could help underserved and rural communities by reducing structural barriers to care and potentially decreasing reliance on emergency departments for non-urgent needs.

The findings highlight the value of integrating AI symptom checkers with provider workflows, as demonstrated by Simtomi-Care. In our data, this integration was reflected in how U.S. providers described this module as "another arm of Simtomi," highlighting its role in bridging the patient-facing app with clinical workflows. This integration was highly valued by professionals in both countries. In Korea, where a much higher proportion of consultations are conducted by specialists than in the U.S.[31], physicians emphasized the value of customized pre-consultation questionnaires rather than reliance on differential diagnosis lists. By contrast, the U.S. context places greater weight on primary care and general practitioners, who prioritize the accuracy of differential diagnosis. This structural difference also aligns with role-specific practices: general practitioners often prioritize broad differential coverage, while specialists need more fine-grained, domain-specific questioning [32]. These findings illustrate why a single AI-driven model may not fully accommodate the diversity of specialties, workflows, and physician preferences. Interviews with Korean clinicians confirmed this demand: while standardized AI-driven symptom questionnaires offered some usability, hospitals preferred customization to adapt items and summary formats to their own institutional practices. Accordingly, the future direction of Simtomi should involve either diversifying models fine-tuned for specific specialties or introducing an administrative service that allows hospitals to directly create and manage their own questionnaires. Such flexibility is expected to enhance the clinical acceptance of Simtomi-Care and serve as a key factor in its broader adoption in Korea. Beyond symptom checking, the Simtomi app also provides a translation function that allows patients to input simple symptoms and communicate their medication needs to pharmacists. Notably, the Simtomi data indicates that pharmacy translation services were used four to five times more frequently than hospital consultations [10]. This suggests that international residents are more likely to seek care through pharmacies rather than clinics, while also experiencing psychological barriers to clinic visits. Several factors contribute to this trend: pharmacies pose lower financial barriers, require shorter interactions, and involve simpler administrative procedures compared to hospitals. Previous studies likewise report that immigrants often postpone or avoid hospital visits due to language difficulties, costs, or lack of insurance, instead relying on self-medication or pharmacists' advice [33], [34]. This context highlights a broader design implication for AI-based health tools: addressing access barriers requires not only linguistic support, but also sensitivity to how care-seeking pathways differ across communities. These extensions have the potential to reduce information gaps and support more informed decision-making in everyday health-seeking situations.

### *B. Limitations and Future Work*

This research has several limitations. The qualitative phases involved relatively small groups of participants; the findings do not capture the full range of perspectives across either healthcare context. The U.S. survey addressed breadth, yet written responses may not provide the same depth as in-person synchronous interviews. Another limitation relates to the context of use. Since the U.S. participants did not evaluate the system in a real clinical environment, the findings should not be interpreted as a direct cross-country comparison of system performance or acceptance. Comparative studies across additional countries would help illuminate how structural differences in health systems shape expectations and perceived value. Future work should therefore examine the system under comparable conditions across healthcare settings. Longitudinal studies with diverse patient groups could provide insights into sustained engagement and the role of digital tools in supporting efficient healthcare access over time.

Appendices: https://drive.google.com/file/d/1R9yYrMK0gT85lKxfyYHnH74gVSlus_iE/view?usp=sharing